\documentclass[10pt,final,journal,letterpaper,twoside,twocolumn]{IEEEtran}

\pdfoutput=1

\usepackage{graphicx}
\usepackage{amsmath}
\usepackage{amssymb}
\usepackage{listings}
\usepackage{color}
\usepackage[hidelinks,bookmarks=false,pdfstartview={XYZ null null 1.00}]{hyperref}
\usepackage{cite}
\usepackage{enumerate}
\usepackage{balance}

\begin{document}

\title{Solar Lab Notebook (SLN): An Ultra-Portable Web-Based System for Heliophysics and High-Security Labs}

\author{Panagiotis~G.~Tsalaportas,~\IEEEmembership{Member,~IEEE},
Vasileios~M.~Kapinas,~\IEEEmembership{Member,~IEEE},\\
and~George~K.~Karagiannidis,~\IEEEmembership{Fellow,~IEEE}
\thanks{P.~G.~Tsalaportas is with the National Aeronautics and Space Administration
       (NASA), Greenbelt, MD 20771 USA (email: ssystemx@gmail.com).}
\thanks{V.~M.~Kapinas is with the Department of Electrical and Computer Engineering,
        Aristotle University of Thessaloniki, 54 124, Thessaloniki, Greece (email: kapinas@auth.gr).}
\thanks{G.~K.~Karagiannidis is with the Department of Electrical and Computer Engineering, Aristotle
        University of Thessaloniki, 54 124, Thessaloniki, Greece, and  with the Department of Electrical
		and Computer Engineering, Khalifa University, PO Box 127788, Abu Dhabi, UAE (email: geokarag@auth.gr).}%
\thanks {This work has been co-financed by the European Union (European Social Fund - ESF) and Greek national
funds through the Operational Program ``Education and Lifelong Learning'' of the National Strategic Reference
Framework (NSRF) - Research Funding Program: THALES-NTUA MIMOSA: Reinforcement of the interdisciplinary
and/or interinstitutional research and innovation.}%
\thanks{Digital Object Identifier $\#\#\#$}%
}



\maketitle

\begin{abstract}
This paper introduces the Solar Lab Notebook (SLN), an electronic lab notebook for improving the process of recording and sharing solar related digital information in an organized manner. SLN is a \textit{pure web-based} application
(available online: \url{http://umbra.nascom.nasa.gov/sln}) that runs client-side only, employing a clean and very friendly graphical user interface design, and thus providing a true cross-platform user experience. Furthermore, SLN leverages unique technologies offered by modern web browsers, such as the FileReader API, the Blob interface and Local Storage mechanism; it is coded entirely using HTML5, CSS3, and JavaScript, and powered by the extremely well documented XML file format. For high-security labs, it can be utilized as an ultra-portable and secure digital notebook solution, since it is functionally self-contained, and does not require any server-side process running on either the local or a remote system. Finally, the W3C XML Schema language is used to define a list of rules, namely a data standard, that an SLN file must conform to in order to be valid. In this way, developers are able to implement their own validation functions in their projects, or use one of the freely available tools to check if a data file is properly structured. Similarly, scientific groups at different labs can easily share information, being confident about the integrity of the exchanged data.
\end{abstract}


\section{Introduction}\label{sec:Intro}

Scientists and engineers have long used laboratory notebooks as a way of recording not
only experimental results, but also for making notes, drawing schematics, making ``back
of the envelope'' calculations, and even, with appropriately translucent adhesive tape,
including samples of experimental materials. Usually, the notebooks contained graph-ruled
paper to ease the making of plots of experimental results. In short, the development of
science and scientific methodology from the ancient civilizations through to the modern era
is strongly connected to the use of paper as a means to facilitate various lab experiments~\cite{Shankar_Recordkeeping}.

Today's scientists and engineers have similar needs, but a much wider range of objects they
would like to record, beyond simply making their own notes. With the advent of personal
computers in laboratory environments and rich digital media from new scientific instruments,
came the need to store, search, combine, extract, and share data, as well as other pieces of
information in an organized manner~\cite{Nature_Coding_Apps,Nature_Quantitative_Data}.
In the context of a laboratory, for example, where scientists investigate the physics of the Sun,
a scientist who is interested in solar activity and its effects on the Earth, might want to
record the uniform resource locators (URLs) of web pages with pertinent information on a given
event or condition, static images of the Sun in various wavelengths sampling different temperature
regimes in the solar atmosphere, digital videos of time series of such images, and images and
time-lapse videos in derived quantities, such as magnetic field strength or line-of-sight velocity~\cite{Zhang_Flare}. As a practitioner of the nascent field of heliophysics, a solar physicist might also want to record the impact on the geospatial environment with plots of in situ measurements of magnetic field, solar wind speed upstream of the Earth, and indices of terrestrial magnetic activity, such as the planetary $K$-index~\cite{NASA_Solar_cycle}. In addition, it might be worthwhile to add links to space weather prediction model runs, so the laboratory researcher could later compare actual storm arrival time and severity with those forecasts.

Given the nearly universally digital nature of the objects a solar physicist or researcher in a similar field
would like to have in his/her notebook, some fundamental inefficiencies of keeping notes in paper-form
were revealed~\cite{Science_The_Paperless_Lab}. Preserving scientific records in a traditional paper
notebook is a time-consuming process that scales linearly with the amount of data to be recorded, while
paper notebooks have also proven to be vulnerable to damage and loss. Hence, taking into account that
information archiving is cardinal to the practice of science, it became clear that the solution of recording
solar data in a \emph{digital notebook} was the most sensible approach.

Well known as electronic lab notebooks (ELNs), digital notebooks have evolved in an effort to tackle
the drawbacks of keeping paper records~\cite{Nature_A_New_Leaf,Nature_The_Digital_Lab}. Offering
standardized ways of storing data and providing ad-hoc mechanisms to categorize information in realtime,
they have set scientists free to focus on more important tasks. Gradually, ELNs have shifted themselves
from complementary tools to indispensable lab mechanisms in many areas of research such as biology,
medicine, materials science, bioinformatics, geoscience, and astronomy~\cite{ELN_microscope,Chem_Soc_Rev_Lab_Notebooks,Hardy_Drug}.

For the projects conducted at the Solar Data Analysis Center (SDAC) of the National Aeronautics and Space
Administration (NASA), current commercial ELN products offered some but not all of the capabilities needed to
support the aforementioned everyday workflow for solar research. What is more, proprietary software did not
offer the advantages of open-source software for improvement and enhancement by an interested user
community. Therefore, work on a type of ELN focusing on solar physics was initiated, resulting in the solar lab
notebook (SLN) described in this paper. SLN was first debuted on May 2013, being operational at the URL
\url{http://umbra.nascom.nasa.gov/sln}. Since then, it has been a powerful asset to the daily workflow of
solar scientists across NASA and European Space Agency (ESA) research centers, with an increasing interest
momentum from other groups.

In different research contexts, various web-based or client software tools have been designed so far for the classification, distribution, and analysis of data acquired by remote sensing satellite systems. Typical examples of such projects related to Earth Observation include the pioneering ``WMPS'' that is a web-based system which allows an inexperienced user to perform unsupervised classification of satellite/airborne images~\cite{IEEE_Select_Topics_System}, the ``Giovanni'' for management of Earth science data~\cite{IEEE_Giovanni}, the ``GeOnAS'' for discovery, retrieval, analysis, and visualization of geospatial data~\cite{Zhao_Geospatial}, and the ``EOSDIS'' testbed system for online access of environmental satellite data~\cite{Emery_weathersat}. Other tools, related to sea and river data monitoring are the ``ARKTOS'' for classification of synthetic aperture radar sea ice images~\cite{TGARS_ARKTOS}, the ``DAMA'' for infrared sea modeling and analysis~\cite{TGARS_DAMA}, the ``SeaWiFS'' for ocean color data archive and distribution~\cite{TGARS_SeaWiFS}, and the ``RivWidth'' for calculation of river
widths from remotely sensed imagery~\cite{TGARS_RivWidth}. In the scientific discipline of time-series analysis, two very indicative tools are the ``TiSeG'' and ``TimeStats'' systems~\cite{TimeStats,TGARS_TiSeG}. Finally, in Heliophysics research, various web-based projects have been developed, such as the ``SolarRad'' for real-time distribution of satellite incident solar radiation~\cite{Tanahashi_SolarRad}, and the ``GSRad'' for estimation of solar radiation~\cite{Donatelli_SolarRad}, among others.

A usage scenario in the domain of remote sensing could commonly be found in the process of keeping notes related to  aps derived by GIS applications. For example, a geoscientist that uses the software products ArcGIS and QGIS could utilize SLN to record personal notes for sections of maps he is working on, along with scaled map images for each record. Moreover, he could associate the corresponding metadata files of his projects with these records, and bundle his notes and data together in a single SLN file for future reference.

\begin{figure*}[t]
  \centering
	\includegraphics[width=\linewidth]{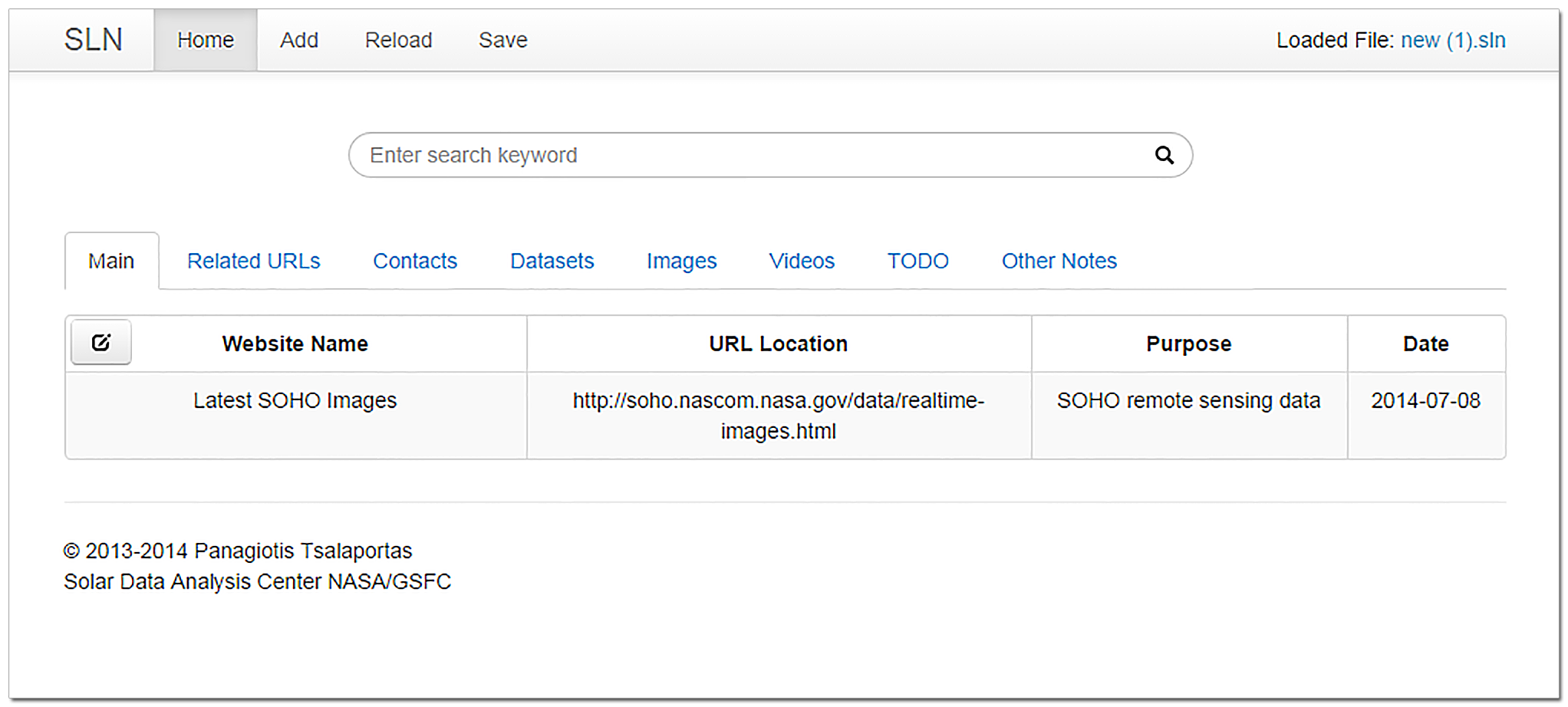}
	\caption{The SLN system's main page. It is designed to have a clean
	             interface with as few buttons as possible, and requires a very small learning
							 curve for new users to start editing. At the top of the window, the interface
							 shows the main menu buttons with which the user can add new entries and
							 export (e.g., save at client-side without connecting to a server) the modified
							 data as a local XML file.}
  \label{fig:SLN_Tab_Main}
\end{figure*}

The rest of the paper is organized as follows: in Sections~\ref{sec:SLN_Motivation} and \ref{sec:Solar_Lab_Notebook},
we discuss the motivation behind the production of SLN in greater detail and probe into the facets of using
modern web technologies to achieve unique product characteristics. Then, Section~\ref{sec:SLN_File_Verification}
scrutinizes the process of validating data integrity, while Section~\ref{sec:Tech_Dev_Considerations} explores the
technical considerations that were taken into account while developing this project.

\begin{figure*}[t]
  \centering
	\includegraphics[width=\linewidth]{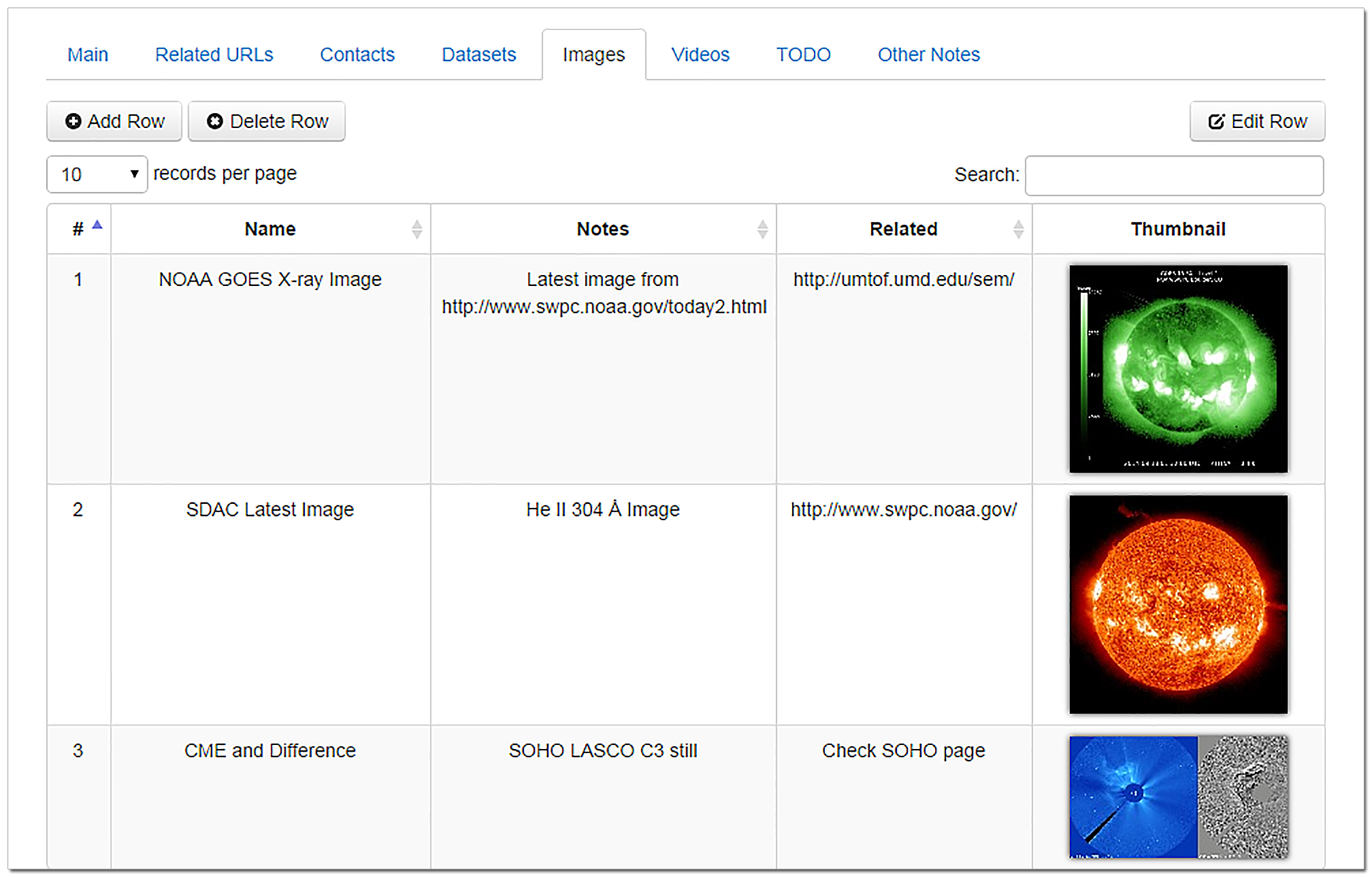}
	\caption{The \textit{Images Tab} is a powerful asset to the work of a solar astronomer. It allows him/her to save and list solar images taken from a variety of sources in a tabular format, from where it is easy to have a direct overview of all available data for a project. Each image entry is associated with relative notes that can be filtered through the tab's search box.}
	\label{fig:SLN_Tab_Images}
\end{figure*}

\section{Motivation for the Development of the Solar Lab Notebook and its Contribution}\label{sec:SLN_Motivation}

\subsection{Desired Project Specifications}

Before exploring the inner workings of this project, it is important to understand the factors that influenced the
motivation of developing SLN, and consequently the reasons on why it has been embraced by so many members of
the scientific community. In particular, the Solar Data Analysis Center at NASA was given the following scenario:

\begin{itemize}
\item A new software tool had to be adopted by the lab's astronomers that could improve the process
of recording and sharing solar related digital information. For this case, an ELN seems to be the most
appropriate option, due to the advantages mentioned in Section~\ref{sec:Intro}. The specialized solar
ELN should allow smooth importing of decades-old surviving handwritten notes and images. Being able
to store notes in a variety of spoken languages would also appeal to a larger user base.
\item It should define a file format standard for data archiving that would be future-proof. Third-party
software applications in the future must be able to extract stored information without having access to
the original source code of this tool or its architecture specifications, and without causing
any data corruption issues.
\item It should be fully functional even in high-security totally isolated lab environments. This translates
to laboratories with no internet connection available, possibly even without local area network (LAN)
connectivity, where users are not allowed to install or run non-regulation approved applications, such as,
for instance, starting a server process.
\item The solar ELN should be able to run on multiple operating systems (OS), while providing the same
user experience. It should also employ a simple graphical user interface (GUI) with a very small learning curve,
given that users usually do not have enough time to invest in new tools. No installation should be required on
any OS, given the security constraints mentioned earlier.
\item The software should be developed using widespread technologies, so as to enable future modifications
and upgrades easily, according to individual research needs.
\end{itemize}

Considering these specifications, currently available products were deemed inapplicable as possible solutions,
since they were not able to simultaneously fulfill all requirements. Hence, it became apparent that a new
custom-built solution had to be developed.

\subsection{Creating a Unique Solution}

To address the concerns and requirements described above, the software in question should focus on the
following main key areas: i) feature a GUI that the scientific audience is familiar with,
and ii) be designed with a substantially different architecture than other commercial ELNs in order to
achieve maximum portability and provide the requested archiving capabilities.

The SLN project has been coded from a clean slate to now offer these unique features:

\textbullet \ \ It employs a clean GUI design and an organization of components that allows for easy access
to information, importing and exporting data (Fig.~\ref{fig:SLN_Tab_Main}). New users are able to pick up
how the application works within minutes, yielding its rapid adoption among members of a lab.

Moreover, it is \textit{platform-independent} and able to run on any OS a browser runs, including Microsoft
Windows, Apple OS X, Linux, *BSD, UNIX systems, etc.~\cite{IBM_Mobile_app}. It also achieves true
\textit{cross-platform user experience} by displaying the GUI through a web browser window. An important
benefit of using a web GUI is that components look and act the same on all systems where a browser is available~\cite{Intel_Cross_Platform_HTML5}. Currently, Google Chrome (and Chromium) offers the full set of the project features, with support for other browsers following up soon.

\textbullet \ \ The software leverages unique technologies offered by modern web browsers. It is coded
entirely using hypertext markup language (HTML), cascading style sheets (CSS), and JavaScript
which represent a very large developer base with numerous documentation materials~\cite{Accenture_HTML5_Dev}.
This allows for future upgrades while keeping developing costs at a minimum. Furthermore, being a pure web
application that runs client-side only, implies zero installation hassles and instant new computer system migrations,
as there is no need to install files on the host system~\cite{IBM_Mobile_app}.

\textbullet \ \ For high-security labs, it can be utilized as an ultra-portable and secure digital notebook
solution. Due to the fact that LANs are susceptible to a variety of man-in-the-middle (MITM) attacks~\cite{IBM_LAN_MITM_List} (e.g., network sniffing, address resolution protocol (ARP) cache poisoning,
dynamic host configuration protocol (DHCP) spoofing, etc.), secure networks follow strict operational
and architectural rules and guidelines, which most often include physical network separation for different
teams, groups, and facilities~\cite{NIST_ICS,CPNI_SCADA,XEROX_MFP,SIEMENS_Security}. In restricted
environments, the centralized client-server model, which most ELN commercial products have, is still
an obstacle for practical deployment to multiple isolated subnets and proves to be challenging for
effective network administration. On the contrary, SLN is functionally self-contained and does not
require any server-side process running on either the local or a remote system. A simple web browser
window is adequate to provide full functionality for importing, editing and extracting information, thanks
to the new capabilities implemented in modern browser versions.

\textbullet \ \ SLN is powered by the extensible markup language (XML) file format to store information,
which is an extremely well documented format~\cite{Cageo_XML_Geosciences}. Also, by using Base64
encoding, it offers the ability to contain images and videos inside a simple text file~\cite{IETF_RFC_2397},
as shown in Fig.~\ref{fig:SLN_Tab_Images}. With dozens of ready-to-use parser libraries, XML is future-proof
for reading data in almost all programming languages available.

\textbullet \ \ Most modern web browsers support displaying characters from both the Unicode basic
and supplementary multilingual plane (BMP and SMP), which means that browsers are able to display
\emph{utf-8} loaded data written by scientists in nearly every modern spoken language in the world
along with hundreds of special characters~\cite{Unicode_Specification}. To put it in perspective, this
includes characters in Linear A, Linear B, Ancient Numeric Systems, Egyptian Hieroglyphs, and many more.

\textbullet \ \ Taking the project a step further, a standard has been defined for storing information
inside SLN files (see Section~\ref{sec:SLN_File_Verification}) that data must conform to in order to be valid. Developers are encouraged to implement their own data validation functions into their software or use one of
the freely available tools to check if a data file is properly structured. Accordingly, scientific groups at different
labs and work-environments can more easily share information and be confident about the integrity of data they
exchange.

\section{The Solar Lab Notebook}\label{sec:Solar_Lab_Notebook}

At the heart of the SLN system lies the entity of a \emph{website}, as the daily workflow of a solar
physicist usually involves browsing a number of different websites that function as gateways to
near-realtime as well as archived information on the Sun's active regions and solar activity. Some well
known gateways are: the Solar Dynamics Observatory (SDO) Image Data webpage~\cite{SDO_Website},
the National Oceanic and Atmospheric Administration (NOAA) Space Weather Prediction Center WSA-ENLIL
Model Prediction link~\cite{NOAA_Website}, the Lockheed Martin Solar and Astrophysics Laboratory
(LMSAL) Latest Events website~\cite{LMSAL_Website}, etc. Thus, based on this common workflow
step, SLN was designed to organize notes in groups (entities) around the different gateways an
astronomer utilizes to find data for his/her work.

\subsection{Sample Workflow Scenario}\label{subsec:SLN_Workflow_Scenario}

Having delineated a general overview of the system's characteristics in Section~\ref{sec:SLN_Motivation},
we can now present a sample workflow scenario that a scientist is likely to follow during his/her everyday activities.
As mentioned earlier, the SLN GUI has evolved to encompass all of its functionalities towards searching and
creating ``website'' groups of data.

After the webpage interface is loaded inside a browser window, a user can either load an existing SLN file through
the ``Load'' button or create a new file through the ``New'' button, located at the top of the page. When an existing file gets loaded, the first website entry is set as active and gets automatically displayed on the Main tab (Fig.~\ref{fig:SLN_Tab_Main}); all data in the other tabs are now related to this website entry. The user can change the current active entry at any time through the main search input, right above the page's tabs. The main search input acts as the cornerstone with which a physicist can seek stored information by typing keywords associated with a ``website'' gateway, and its functionality resembles the \textit{instant search} feature of web search engines, updating results for every subsequent character the user enters.\footnote{We note here that the search functionality in SLN is powered by a module of the Bootstrap web framework, which has been expanded and customized for use with the specific type of data utilized in this application, having the entries index generated ad-hoc in memory.}

\begin{figure}[t]
  \centering
	\includegraphics[width=\columnwidth]{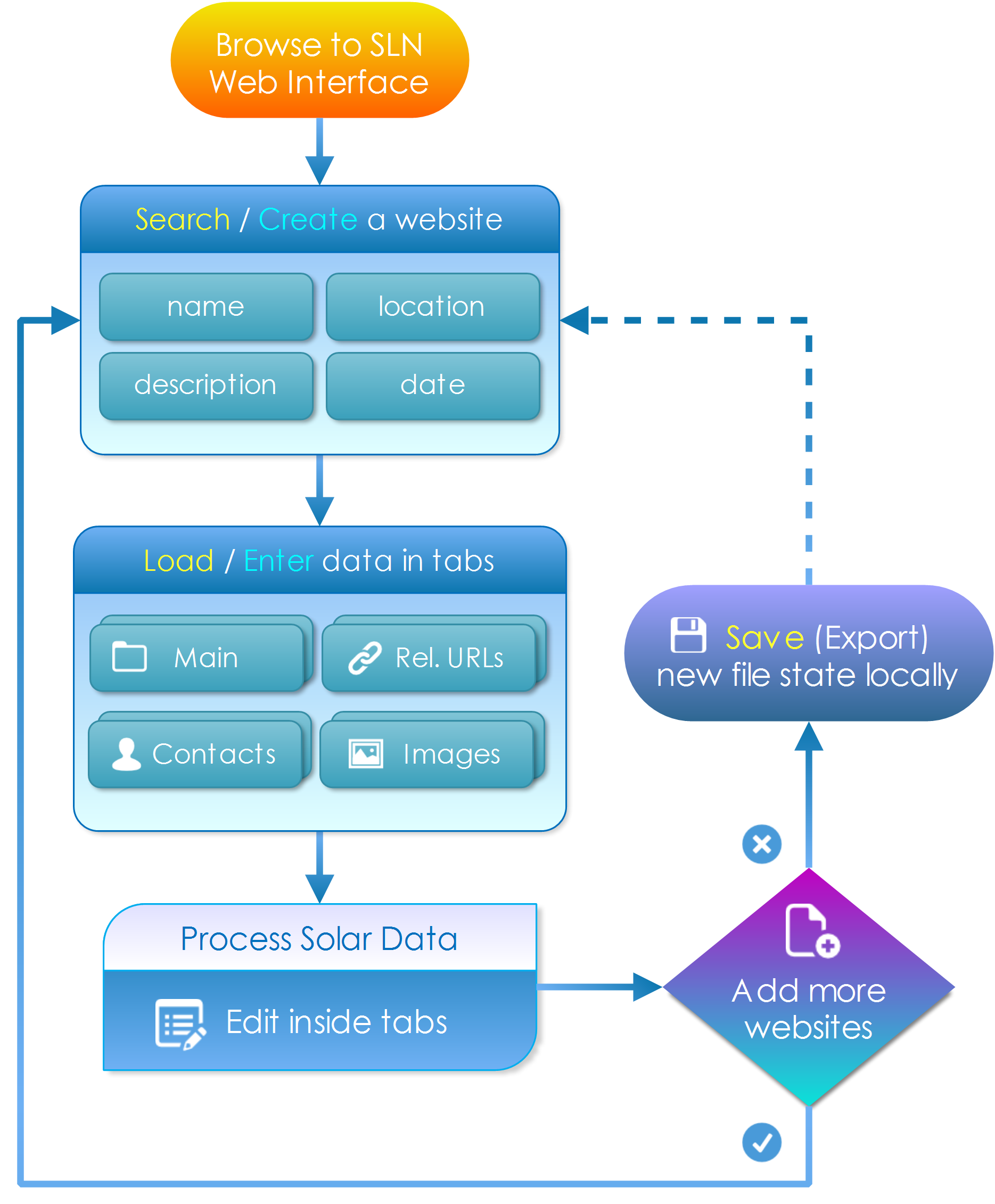}
	\caption{Flowchart describing the usage of the SLN web-based interface. A user can
	             practically export anytime the current state of the file he/she is editing. Note that a dashed
							 line indicates an optional step.}
  \label{fig:SLN_Workflow}
\end{figure}

Clicking on the button located on the left of the ``Website Name'' column of the Main tab (Fig.~\ref{fig:SLN_Tab_Main}), will allow the user to edit the parameters of this entry, such as its
\textit{URL Location}, the \textit{Purpose} of recording the entry, as well as the \textit{Date} that this
website was visited and found to have important scientific data. Subsequently, the user will probably also
click on the other tabs to log/edit new information, as described in the next subsection. In general, the
workflow of SLN can be summarized in Fig.~\ref{fig:SLN_Workflow}.

After the process of recording and editing data is complete, it is easy to export them as a bundle in a new SLN file
by clicking on the ``Save'' button, which will generate a local SLN file. In order to share this data, a scientist can add the SLN file as an attachment to an email, store it in a server location for remote access, or transfer it through a portable storage device.

\subsection{Application Web-Based Interface}\label{subsec:SLN_Web_GUI}
As Fig.~\ref{fig:SLN_Tab_Images} illustrates, the GUI has an overall clean and easy to use design that presents
a simple tab layout to navigate users through a variety of sections. There are eight sections/tabs, each specialized
for a different kind of note keeping. All records, either text or images, reside in a table format, while table entries can be easily added, edited or deleted through the tab's ``Add Row'', ``Delete Row'' and ``Edit Row'' buttons. On every tab, an exclusive search field is available for its table that allows for easy sorting and filtering of information.

\emph{\textbf{1) Main Tab:}} The Main tab page, shown in Fig.~\ref{fig:SLN_Tab_Main}, displays the
``website'' entry which is currently active. All other tabs-sections in the browser window will display
information related to this entity.

\emph{\textbf{2) Related URLs Tab:}} On the Related URLs tab, the user can save URLs of
websites that are relevant to the active ``website'' entry. For example, a solar data gateway may provide
further data about an event with links to other gateways. These links can be easily included in this tab for
direct access to the external source of data.

\emph{\textbf{3) Contacts Tab:}} Similar to the previous tab, the Contacts section, acts as a
list of contacts that the solar scientist may wish to communicate with or are relevant to a current project
he or she is working on.

\emph{\textbf{4) Datasets Tab:}} Under the tab of Datasets, the user can store large amounts of \textit{text-form}
raw data organized as table entries. With each entry there is a button that when clicked will open a new
browser window displaying the entry's data; this data may also be saved as files locally in the same way someone
would save an HTML page from a website. Some sample entries could be: full HTML/XML documents, \mbox{utf-8}
text files (e.g., CSV), VOTable tabular data, any XML-based files (e.g., SVG, RSS, KML, Atom, X3D), Raw text-data
output from scientific instruments, other SLN files, etc.

Regarding \textit{binary} file formats, such as HDF, it should be important to clarify that HDF files (as well as netCDF) are not in textual form, and special tools are needed to parse byte ranges of data organized in the different sections of a file. On the contrary, textual data are easier to parse within browsers, and there are dozens of readily available parser libraries to be used for software projects. Future versions, though, of the SLN application may also allow scientists to embed binary file formats like PDF (\textit{*.pdf}), HDF (\textit{*.hdf}), netCDF (\textit{*.nc,*.cdl}), and Word (\textit{*.docx}) documents.

There is no preset size limit for text files or data that can be stored inside this table. It is only the browser's capabilities and the available free system RAM that can impose limits on the amount of data to be stored. Environments with better hardware specifications, e.g., more free memory and a faster processor, will perform accordingly better. On several internal benchmarks performed using Chrome while developing this software, the application was able to process SLN files of over $400$ Megabytes (MB) in size containing datasets and images, after a loading phase of $3-4$ seconds. Since data loaded from the files reside in the browser's process RAM, it is advised for a computing environment to have at least $200$ MB of free memory while the application is active.

\emph{\textbf{5) Images Tab:}} The Images tab (Fig.~\ref{fig:SLN_Tab_Images}) is one of the most important
sections utilized by solar scientists because it allows them to add entire solar images (or portions of them) to an
SLN container file. Such an aspect plays a major role in Heliophysics research by providing, for example, the opportunity to display photos at different wavelengths derived from different solar gateways in sortable table columns, so that the physicist may have an immediate overview of all available data for an event. Illustrated in
Fig.~\ref{fig:SLN_Tab_Images}, every entry (row) has a given symbolic \textit{Name}, some \textit{Notes},
possibly a directly \textit{Related URL} of a webpage, and a photo \textit{Thumbnail}. Clicking on the thumbnail
opens a new window with its full scale image that may also be saved (extracted) as a normal image file on the
user's computer. In order to add new photos to the table, SLN provides a unique image editing environment, the
Image Editor.

\begin{figure}[t]
  \centering
	\includegraphics[width=\columnwidth]{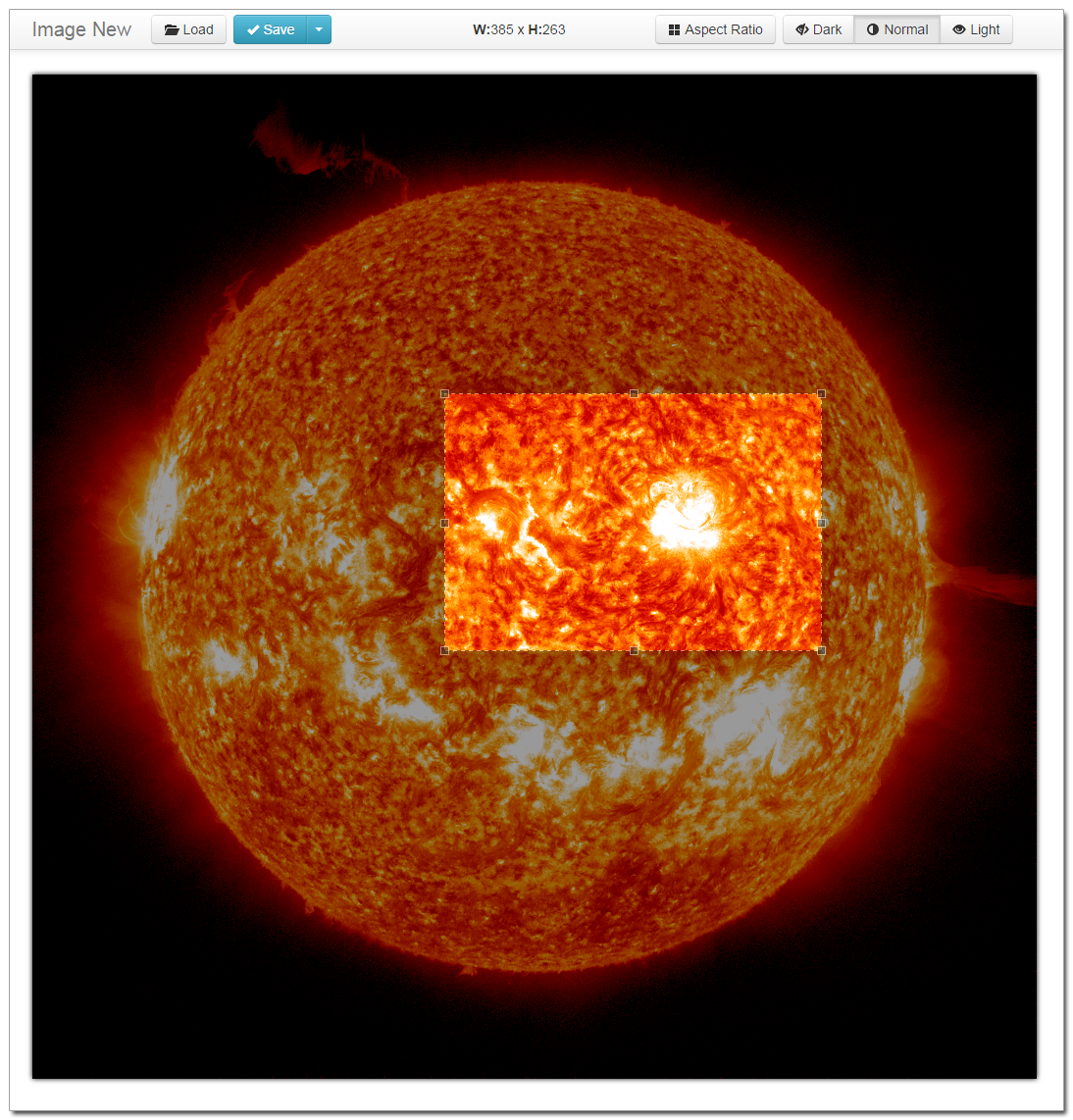}
	\caption{An example demonstrating the SLN Image Editor running in a separate
	             window from the main application. Apart from being able to save the entire image in full
							 size or in smaller scale, the user has also the ability to focus and save a free-form
							 rectangle of the photo that better suits the needs of his/her projects. This eliminates
							 the need to use an external photo editor and adds to the improvement of the daily workflow.}
  \label{fig:SLN_Image_Editor}
\end{figure}

Launching the embedded Image Editor of SLN opens a new browser window. The Editor offers an easy way for
astronomers to load a solar photo from their desktop and store it to an SLN file in full size (up to $2048 \times 2048$ pixels) or in scales of \textit{1:2}, \textit{1:4}, and \textit{1:8} of its original dimensions. Furthermore, a scientist may select to keep only a portion of the image that is more relevant to his/her work, as demonstrated in
Fig.~\ref{fig:SLN_Image_Editor}. A special feature of the SLN software is that for every full size image a user chooses to store, a smaller extra thumbnail is automatically created and it is then the one that gets displayed in the tab's table; this mechanism greatly increases the responsiveness of the application because when the browser process renders the table it simply has to place the saved thumbnail in the correct row instead of also having to calculate on-the-fly
thumbnails for full-size photos.

Supported image formats by the Image Editor are mentioned in Table~\ref{tab:SLN_Image_Formats}. Uploaded images
must adhere to this table of support according to each browser, since SLN utilizes the browser's native image display
functions.

\emph{\textbf{6) Videos Tab:}} With the Videos tab, the solar physicist will have the ability to save time-lapse videos of solar activity, like media files provided from solar gateways. Although the SLN architecture supports such features, the current state of browsers does not yet offer this functionality, at least in an adequately responsive way. Wide adoption of open video codecs must first be implemented in browsers (e.g., \textit{webm}), and it is expected that future browser versions will allow native editing of video files. When these technologies become more widely offered, this tab will be ready to function normally.

\emph{\textbf{7) TODO Tab:}} The purpose of the TO-DO tab, as the name suggests, is to act as a reminder list for things a scientist has scheduled or would like to do. A sample entry for a solar astronomer would be to check the National Oceanic and Atmospheric Administration (NOAA) Solar Plots page~\cite{NOAA_Website} on a specific date. By having this to-do note associated with the current active ``website'', it becomes easier to organize the everyday workflow for imminent actions or future planning.

\begin{figure}[t]
  \centering
	\includegraphics[width=\columnwidth]{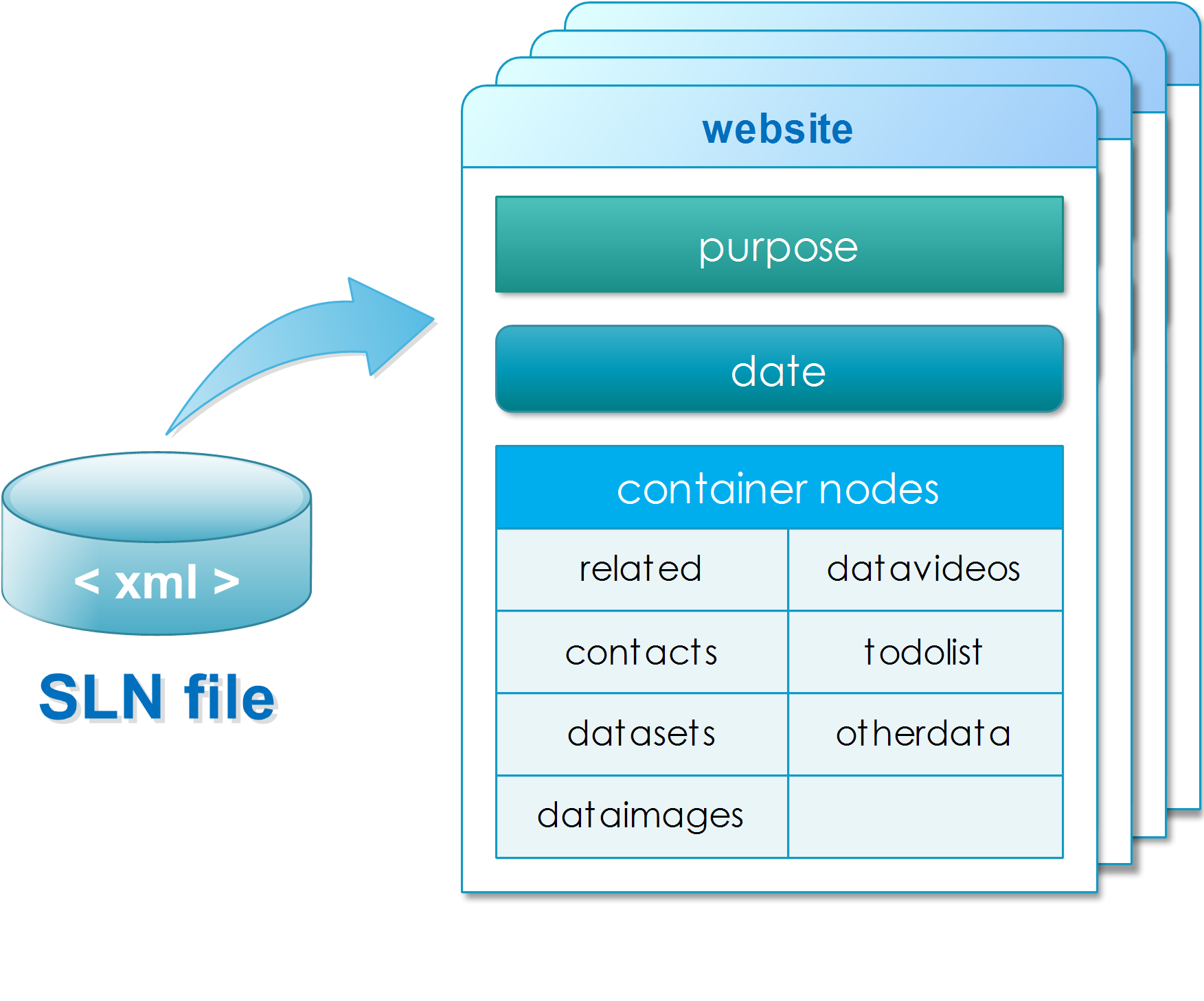}
	\caption{Data inside SLN files are saved in XML format. Every file acts as a container of
	             data and has multiple \lstinline[basicstyle=\ttfamily\footnotesize]|website| elements,
							 each having two single nodes, \lstinline[basicstyle=\ttfamily\footnotesize]|purpose| and
							 \lstinline[basicstyle=\ttfamily\footnotesize]|date|, and a number of container nodes.}
  \label{fig:SLN_File_Structure}
\end{figure}

\begin{table}[t]
  \centering
  \caption{Solar Lab Notebook Supported Image Formats}
  \label{tab:SLN_Image_Formats}
  \begin{tabular}{p{3cm} p{5cm}}
  \hline \hline
	\\ [-0.15cm]
  \multicolumn{1}{c}{\textbf{Browser}} & \multicolumn{1}{c}{\textbf{Image Format}} \\ [0.1cm]
	\hline
	\\ [-0.15cm]
	\multicolumn{1}{c}{Apple Safari} & \hspace{0.35cm}JPEG, GIF, PNG, SVG, TIFF,  BMP \\ [0.2cm]
  \multicolumn{1}{c}{Mozilla Firefox} & \hspace{0.35cm}JPEG, GIF, PNG, SVG, APNG, BMP \\ [0.2cm]
  \multicolumn{1}{c}{Google Chrome} & \hspace{0.35cm}JPEG, GIF, PNG, SVG, WEBP, BMP \\ [0.15cm]
	\hline \hline
	\end{tabular}
\end{table}

\emph{\textbf{8) Other Notes Tab:}} Selecting the Other Notes tab, a scientist can keep various kind
of text notes that do not fall under any of the previous tab categories. This tab often serves as a temporary digital
space for simple notes, allowing data in the other tabs to not be mixed with different kinds of entries thus keeping a
clean taxonomy. Of course, it may also serve as a general note-keeping utility.

\definecolor{green}{rgb}{0,0.6,0}
\definecolor{darkblue}{rgb}{0.0,0.0,0.6}
\definecolor{orange}{rgb}{1,0.27,0}
\definecolor{cyan}{rgb}{0.0,0.6,0.6}
\lstdefinelanguage{XMLSLN} {
  keepspaces=true,
  frame=single,
  tabsize=2,
  showtabs=false,
  showspaces=false,
  showstringspaces=false,
  breakatwhitespace=false,
	basicstyle=\ttfamily\scriptsize,
	commentstyle=\color{green},
  identifierstyle=\color{darkblue},
  keywordstyle=\color{cyan},
  keywordstyle=\color{orange},
	morestring=[s]{"}{"},
  morecomment=[s]{?}{?},
  morecomment=[s]{!--}{--},
  commentstyle=\color{green},
  moredelim=[s][\color{black}]{>}{<},
  moredelim=[s][\color{orange}]{\	 }{=},
  stringstyle=\color{cyan},
  identifierstyle=\color{blue},
	escapechar=\&,
	alsoletter=:,
  morekeywords={
	  xmlns, xmlns:xsd, xsi, xmlns:xsi, schemaLocation, name,
		surname, location, value, elementFormDefault, use,
		attributeFormDefault, targetNamespace, maxOccurs, type}
}

\begin{figure}[t]
\begin{lstlisting}[language=XMLSLN]
<?xml version="1.0" encoding="utf-8"?>
<sln xmlns="http://umbra.nascom.nasa.gov/"
     xmlns:xsi="http://www.w3.org/2001/XMLSchema-instance"
	   xsi:schemaLocation="http://umbra.nascom.nasa.gov/ &$\hookleftarrow$&
		  http://umbra.nascom.nasa.gov/sln/schema/sln.xsd">

<website name="Latest SOHO Images"
         location="http://soho.nascom.nasa.gov/data/&$\hookleftarrow$&
				            realtime-images.html">
  <purpose>SOHO remote sensing data</purpose>
  <date>2014-09-05</date>
  <related>
    <reluri value="http://sdo.gsfc.nasa.gov/data/">
      <notes>SDO near-realtime image data</notes>
    </reluri>
    <reluri value="http://soho.nascom.nasa.gov/data/&$\hookleftarrow$&
		                realtime/mpeg/">
      <notes>Near-realtime SOHO MPEG movies</notes>
    </reluri>
    <reluri value="http://www.swpc.noaa.gov/wsa-enlil/">
      <notes>NOAA SWPC WSA-ENLIL Model Prediction</notes>
    </reluri>
    <reluri value="http://www.swpc.noaa.gov/today2.html">
      <notes>Integrated Solar Soft X-Ray flux and &$\hookleftarrow$&
			        satellite environment plots</notes>
    </reluri>
  </related>
  <contacts>
    <contact name="New" surname="Contact">
      <email>email@nasa.gov</email>
      <webpage>nasa.gov</webpage>
      <notes>New Contact</notes>
    </contact>
  </contacts>
	...
</website>
\end{lstlisting}
	\caption{XML architecture of an SLN file. The code above is in accordance to the structure
	             illustrated in Fig.~\ref{fig:SLN_File_Structure}. The rules that set the range of allowed
							 values for each element are described in Fig.~\ref{fig:SLN_Schema}.}
  \label{fig:SLN_XML_Structure}
\end{figure}

\subsection{Underlying XML Architecture}

For the variety of reasons stated in Section~\ref{sec:SLN_Motivation}, the storage mechanism is
orchestrated to function by saving data in plain XML text files instead of using a database. In a sense,
the underlying XML architecture that powers the project's internal storage mechanism poses as a reflection
of the Graphical User Interface. Consequently, information inside SLN files is stored in XML according to a
set of rules imposed by the SLN Schema standard described in Section~\ref{sec:SLN_File_Verification}. Every
SLN file structure contains one or more \lstinline[basicstyle=\ttfamily\small]|website| elements that act as
large containers of information. Following the GUI layout, all \lstinline[basicstyle=\ttfamily\small]|website|
related data are stored in groups (child elements) of Related URLs, Contacts, Datasets, Images, Videos,
TODO notes, and Other solar related notes, as shown in Fig~\ref{fig:SLN_File_Structure}.

A distinctive way of comparing the above model for saving data would be to check the mechanism of storing information inside ZIP archives. In a ZIP container file there might exist a number of folders, each of them containing a number of individual text and image files. Similarly, an SLN file acts as a container of a number of folders (``website'' entries) with stored data and images that a scientist may have selected to record. Since SLN files are created according to their XML Schema standard, the data they contain are structured in a tree-like hierarchy, based on nodes.

To better illustrate this hierarchy, consider the following scenario: when a user adds a new Contact through the application, a \lstinline[basicstyle=\ttfamily\small]|contact| node is generated under the node \lstinline[basicstyle=\ttfamily\small]|contacts|; the name and surname values of the Contact will be stored inside the file as the XML attributes \lstinline[basicstyle=\ttfamily\small]|name| and \lstinline[basicstyle=\ttfamily\small]|surname| (Fig.~\ref{fig:SLN_XML_Structure}). Further details may be stored in the child elements \lstinline[basicstyle=\ttfamily\small]|email|, \lstinline[basicstyle=\ttfamily\small]|webpage|, and \lstinline[basicstyle=\ttfamily\small]|notes|. Thus, by harnessing the power of XML tree structure and attributes, SLN files are organized as \textit{well-defined} documents of data with series of hierarchical nodes.

\section{File Integrity Verification}\label{sec:SLN_File_Verification}

This section aims to take a deeper look at the development of preservation efforts for future-proofing SLN's file format. The ever increasing amount of data in XML format across the globe has long created the need to better describe the content of data structures and constraints in XML files. Thereupon, in order to tackle this impediment several XML schema languages have been proposed throughout the years.

\begin{figure}[t]
\begin{lstlisting}[language=XMLSLN]
<?xml version="1.0" encoding="utf-8"?>
<xsd:schema xmlns:xsd="http://www.w3.org/2001/XMLSchema"
            xmlns="http://umbra.nascom.nasa.gov/"
            elementFormDefault="qualified"
						attributeFormDefault="unqualified"
						targetNamespace="http://umbra.nascom.nasa.gov/">
...
<xsd:complexType name="ContactsType">
  <xsd:sequence>
    <xsd:element name="contact" maxOccurs="unbounded">
      <xsd:complexType>
	      <xsd:sequence>
          <xsd:element name="email" type="xsd:string" />
          <xsd:element name="webpage" type="xsd:string" />
          <xsd:element name="notes" type="xsd:string" />
        </xsd:sequence>
        <xsd:attribute use="required" name="name"
				               type="attribStringType" />
        <xsd:attribute use="required" name="surname"
				               type="attribStringType" />
      </xsd:complexType>
    </xsd:element>
  </xsd:sequence>
</xsd:complexType>
\end{lstlisting}
	\caption{Sample code of the SLN W3C XML Schema. Here, the rules define
	             how a \lstinline[basicstyle=\ttfamily\footnotesize]|contact| element in an SLN file
							 should be structured. It must be noticed that the order of elements inside
							 \lstinline[basicstyle=\ttfamily\footnotesize]|xsd:complexType| and
							 \lstinline[basicstyle=\ttfamily\footnotesize]|xsd:sequence| nodes is
							 important for the validation process.}
  \label{fig:SLN_Schema}
\end{figure}

An XML \textit{schema} language is a precise description of another XML-based language in terms of
constraints on its elements and their attributes~\cite{Murata_XML}.  Take, for example, an SVG image file.
Scalable Vector Graphics, or SVG, is an XML-based vector image format. Opening an SVG file with a text
editor would reveal a hierarchical structure similar to Fig.~\ref{fig:SLN_XML_Structure}. The rules that define
\emph{how} an SVG file should be structured, like what the order of elements should be and what attributes
they should have, are stated in the SVG Schema standard that is similar to Fig.~\ref{fig:SLN_Schema}. By
definition, an SVG file must follow precisely the rules of its standard in order to be valid. Among the most
frequently used schema types are DTD, W3C XML Schema, RELAX NG, and Schematron. The W3C XML Schema
(XSD)~\cite{W3C_XML_Schema} has been selected to describe the SLN data format~\cite{SLN_XSD_Schema}
due to the great number of open source or freely available tools on the market, like the XML C Parser and Toolkit~\cite{XML_Validation_Tool}. Other well known examples of XML Schemas include definitions for the formats:
XAML, KML, RDF, Atom, RSS, COLLADA, MARCXML, MathML, NeuroML, OpenDocument (ODF), PMML, Protein
Data Bank (PDB), X3D, etc.

\textit{Validation} is the process of testing an XML document to check whether it conforms to a schema. When such a validation is true, we can be sure that the stored data follow the desired rules we have set~\cite{Murata_XML}. Defining rules for SLN files has been a lengthy process and one that had to be thoroughly considered with a view of achieving widespread adoption of the project's data format. Frequent feedback from the scientific community has assisted to shape a standard that allows for much freedom regarding values of data structures and attributes, at least towards the use of SLN files in a browser environment. Hence, the SLN Schema (Fig.~\ref{fig:SLN_Schema}) is primarily used to ensure that data are well defined focusing on structure, rather than the range of permitted values for elements.

\section{Technical Development Considerations}\label{sec:Tech_Dev_Considerations}

\subsection{Modern Web Technologies}

The development of the Solar Lab Notebook was based on new technologies offered by browsers, like HTML5,
that make it easier for developers to interact with files, graphics and memory storage mechanisms. It would
therefore be worth presenting some examples of how these technologies could allow the creation of similar
projects for a variety of other areas of research.

First of all, in order for a web application to interact with local files on a client's machine without the need of a
server-side process (local or remote), it is important to communicate with the browser's file access rights in a standardized way. Such is the purpose of the HTML5 FileReader API~\cite{SLN_HTML5_FileReader}, which is basically a method of reading the contents of a file into memory, and which has only recently been available by browser vendors
(Chrome 32+, Firefox 33+, Safari 7.1+). The FileReader interface can be used to asynchronously read a file through
familiar JavaScript event handling. In our project, this API has been the cornerstone of allowing users to ``upload''
local SLN and image files directly to the application's memory space through the client's browser window, thus avoiding the need to ``send'' this data to a server for processing.

In addition to the above, after the user edits some data and adds images to his/her electronic notebook, it
must be possible to save this data without their browser sending a request to a host machine. This is where
the HTML5 Blob interface can become useful to developers~\cite{SLN_HTML5_FileReader}, as it represents file-like
objects of immutable, raw data. This interface has been leveraged in our project, in addition to some other HTML5
mechanisms, towards instantly generating a new SLN file in RAM that can be downloaded locally on a computer.

A key requirement for an application that handles images, is to be able to read graphics at pixel level. Such is the
ability provided by the HTML5 Canvas technology~\cite{SLN_HTML_Canvas}. Recent browser versions incorporate
the new Canvas Blend modes, extending the older basic canvas support, which has been the foundation of developing
the SLN Image Editor. The Editor uses multiple canvas elements in overlay mode for selecting interactively only a portion of a solar image to keep, as well as different canvas elements that perform on-the-fly scaling of the image for the creation of thumbnails creation.

Another new feature utilized by SLN is the usage of HTML Local Storage~\cite{SLN_HTML_Storage}, with which web
applications can store data locally in the user's browser. Before HTML5, application data had to be stored in ``cookies'', included in every request to a server. Local storage does not need to communicate with a server, making it more secure, and can store adequate amounts of data (at least 10MB) without affecting website performance. SLN uses local storage function calls to transfer data between the Image Editor and the main application window. Having multiple intercommunicating windows on a single web application, is something that was previously encountered only in desktop
software solutions.

\subsection{Browser Differences}

When creating web-based applications, developers should be aware of various browser differences regarding the
implementation of standards, as well as the fact that not every browser has implemented all of the new HTML5
technologies. This applies to both the functional aspects of the application as well as the aspects of providing a
consistent user interface (UI). Extensive cross-browser testing is required to ensure that the application has the
same user experience on a browser for all Operating Systems.

An indicative case where specific actions had to be taken towards this goal, was working with the mechanism of
transferring data between separate browser windows though the local storage interface. Although the interface
provides a standard JavaScript API that can be used to detect when new data has been generated with the local
storage functions, it became apparent that the corresponding mechanisms for these events were not functioning consistently between different browsers. For this reason, a custom event-checking JavaScript function was
developed, that would automatically inform the main application procedures when new data was ready to be
transferred from the Editor to the main window. Additional checks had also to be implemented so as to reset this
mechanism if the Image Editor was closed by a user action other than the embedded menu options.

More focus might additionally be needed when working with the new CSS3 standards, affecting directly the appearance of the user interface. Sometimes, specific rules have to be added in style sheets so as to adjust various visual differences between browsers. For example, in SLN, the exact placement of a number of elements on the screen gets modified based on the browser vendor, with an ultimate purpose of giving a more appealing user experience.

Furthermore, each browser's rendering efficiency should be taken into account before releasing the product.
For example, certain SLN functions have been monitored with a memory (RAM) and processor (CPU) profiler to
render the webpage in less than $16$ milliseconds (ms), so that all animations will render smoothly. A well known factor about smooth rendering in computer screens, like a webpage, is that for the common $60$ frames per second (fps)
to be displayed on a screen, there are only $1000/60 = 16.6$ms available to the application in order to calculate
and draw the scene on that frame. This process runs $60$ times every second, and is dependent on what a web
application does on each frame. To provide the best possible experience for users, the developer has to
monitor critical calculation-intensive parts of the code and make adjustments accordingly.

\subsection{Project Security Aspects}

As mentioned in Section~\ref{sec:SLN_Motivation}, at the core of the Solar Lab Notebook design lies the
requirement for a project that can function properly in a possibly isolated environment. This effectively means
that a researcher should be able to record data notes and images derived from scientific instruments, even
when his/her system might not have any kind of connectivity with other computers in the lab.

Towards fulfilling this specification, while not using external encryption libraries for easier maintenance and
extensibility, this application makes prominent use of modern web technologies in order to prevent data
communication between the user's computer and a server. Leveraging HTML5 standards mentioned earlier, SLN
files can be loaded, viewed, edited and generated locally and entirely on the client's browser. Since there is no server process running on a local or host machine, the administrators are not required to take extra measures of security, which would mean more maintenance steps for their network.

Finally, moving the project and existing SLN files from one system to another can be performed directly by
simply transferring them files through a variety of ways. On such occasions, several other centrally managed ELNs demand lengthly procedures of exporting and importing. Instead, in the case of SLN, if a new machine is added to the laboratory, it will be easy for scientists and engineers to migrate to the new system.

\section{Conclusion}\label{sec:Conclusion}

This paper has described the SLN project, namely, the design, development, and deployment of an ultra-portable
and high-security web-based application, aimed to assist research and facilitate collaboration among scientists and
engineers worldwide in heliophysics and space weather forecasts. The SLN tool is a purely scientifically-driven effort
to help the international solar community with the process of recording, combining, and sharing solar related digital
information in an organized and standardized manner. It has been designed to meet all the requirements and
architectural challenges set in Section~\ref{sec:SLN_Motivation} by focusing on four key areas: portability, security,
interoperability, and adaptability. As a result, for some time now, SLN improves the everyday workflow of solar
scientists across different research centers and individuals who are working or are interested in related topics.
Furthermore, the SLN concept has the potential to achieve a high degree of acceptance as well from users specializing
in other disciplines of research and technology. For instance, scientists working in biology or materials science can
fine-tune with minimal changes the SLN W3C Schema standard rules in order to record data related to their own area
of research.

Future evolutions of the SLN software may be expected to demonstrate several more features, including the automatic
fetching of images from certain solar gateways and the support for binary types of data. In this way, we hope that SLN will prove itself to be a beneficial addition to existing software tools for the study of the Sun and its interaction with the Earth's atmosphere.

\section*{Acknowledgment}

The authors would like to especially thank Dr.~Joseph~B.~Gurman from the Solar Data Analysis Center at NASA/GSFC
for his remarkable support and guidance during the development of the project, which made this study possible. In addition, we would like to thank Dr.~Rigas Ioannides from the Radio Navigation Systems and Techniques Section at ESA/ESTEC, for his helpful feedback as well as advice regarding the application features.

\bibliographystyle{IEEEtran}

\vspace{-4.0em}\begin{IEEEbiography}[{\includegraphics[height=1.25in,keepaspectratio]{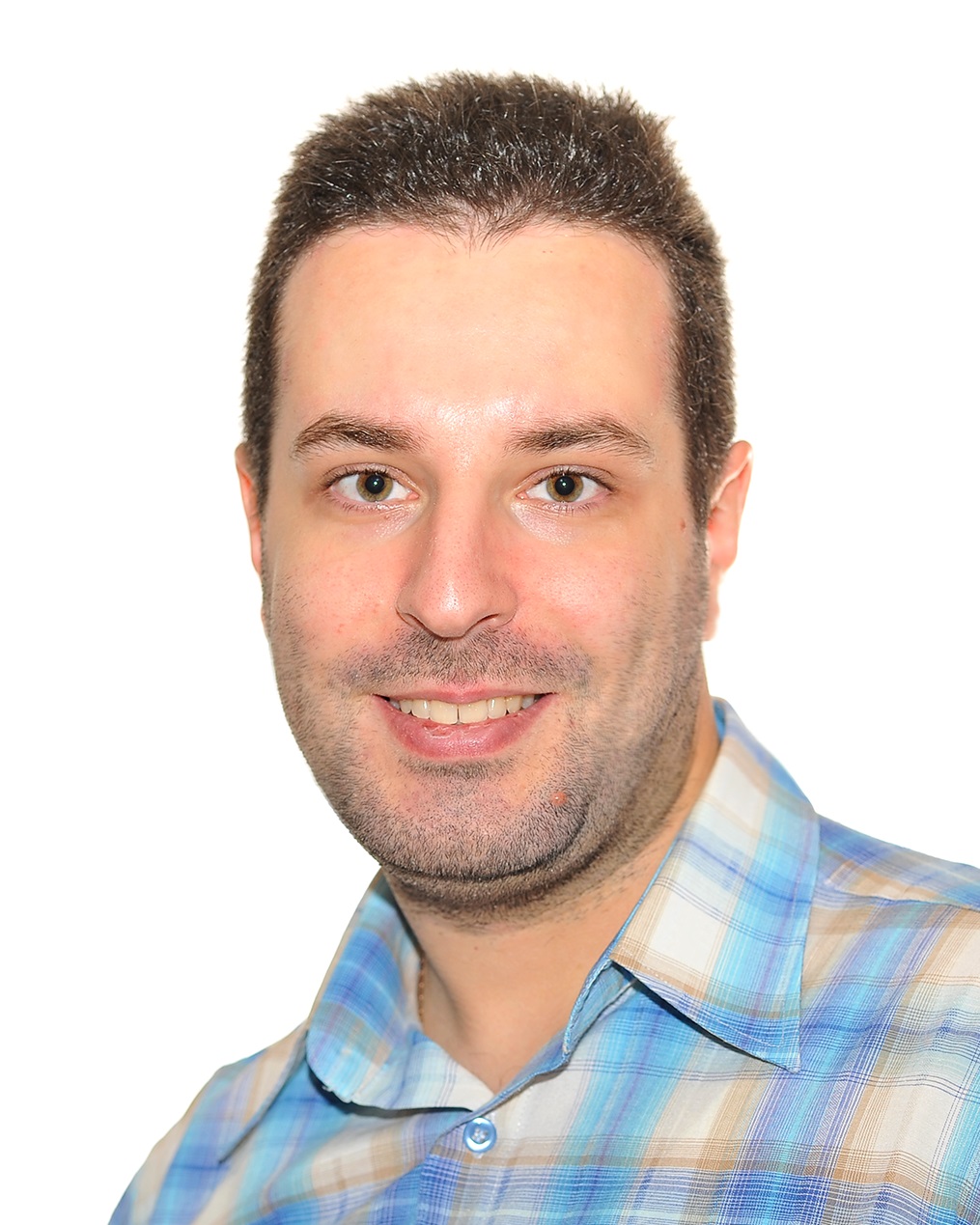}}]
{Panagiotis G. Tsalaportas} (S'10--M'13) was born in Larissa, Greece. He received the B.Sc.-M.Eng. (Engineering Diploma) degree in electrical and computer engineering from the Aristotle University of Thessaloniki (AUTH), Greece, in 2013.

He is currently working as a Research Associate in Software Engineering for the Solar Data Analysis Center (SDAC) at the Goddard Space Flight Center (GSFC), National Aeronautics and Space Administration (NASA). He has been with the NASA SDAC since 2012, designing and developing software solutions towards improving the process of studying the physics of
the Sun. In the past, he was with the Institute for Astronomy, Astrophysics, Space Applications and Remote Sensing (IAASARS), National Observatory of Athens as a research intern. In 2011, he was included in the official list of contributors to the Mozilla Firefox Web browser and since then he has been assisting the development of new features for future Mozilla technologies and products. His current interests are focused on software engineering, space systems and applications, as well as modern Web browser technologies.
\end{IEEEbiography}\vspace{-4.0em}

\begin{IEEEbiography}[{\includegraphics[height=1.25in,clip,keepaspectratio]{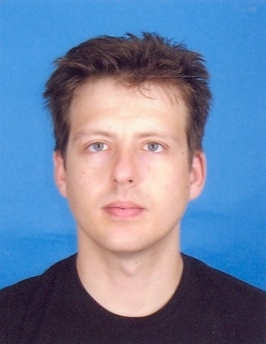}}]
{Vasileios M. Kapinas} (S'07--M'09) was born in Thessaloniki, Greece. He received the Diploma and Ph.D. degrees in electrical and computer engineering from the Aristotle University of
Thessaloniki (AUTH), Thessaloniki, Greece, in 2000 and 2014, respectively. From 2001 to 2005 he was with AUTH as a Graduate Research Assistant at the Signal Processing and Biomedical Technology Unit. During the periods 2006-2009 and 2011-2013 he worked as a Research and Teaching Assistant at the Telecommunications Systems and Networks Lab of AUTH. From 2009 to 2010 he was with the TT\&C Systems and Techniques Section of the RF Payload Systems Division, European Space Research and Technology Centre (ESTEC), European Space Agency (ESA), The Netherlands.

Dr. Kapinas is currently a Postdoctoral Researcher with the Wireless Communications Systems Group, AUTH, focusing on next generation mobile and satellite communication systems. He is also a full member of the AIAA.
\end{IEEEbiography}\vspace{-4.0em}

\begin{IEEEbiography}[{\includegraphics[height=1.25in,clip,keepaspectratio]{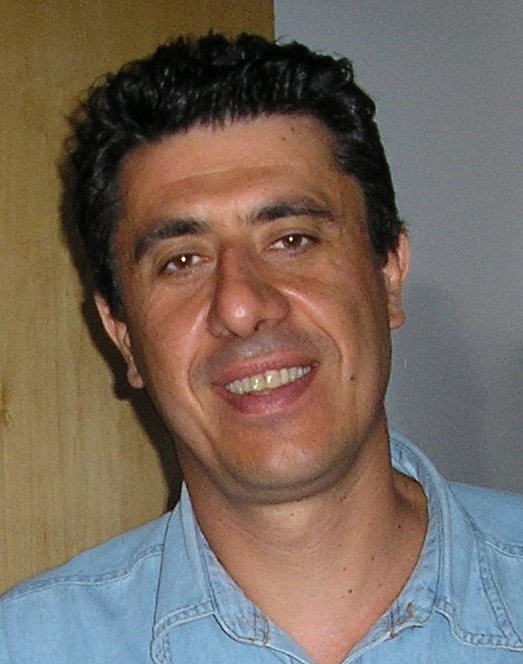}}]
{George K. Karagiannidis} (M'96--SM'03--F'14) was born in Pithagorion, Samos Island, Greece. He received the University Diploma (5 years) and PhD degree, both in electrical and computer engineering from the University of Patras, in 1987 and 1999, respectively. From 2000 to 2004, he was a Senior Researcher at the Institute for Space Applications and Remote Sensing, National Observatory of Athens, Greece. In June 2004, he joined the faculty of Aristotle University of Thessaloniki, Greece where he is currently Professor in the Electrical \& Computer Engineering Dept. and Director of Digital Telecommunications Systems and Networks Laboratory. In January 2014, he joined Khalifa University, UAE, where is currently Professor in the Electrical \& Computer Engineering Dept. and Coordinator of the ICT Cluster. His research interests are in the broad area of digital communications systems with emphasis on communications theory,
energy efficient MIMO and cooperative communications, satellite communications, cognitive radio, smart grid and optical wireless communications. He is the author or co-author of more than 250 technical papers published in scientific journals and presented at international conferences. He is also a co-author of three chapters in books, author of the Greek edition of a book on ``Telecommunications Systems'', and co-author of the book \textit{Advanced Optical Wireless Communications Systems}, Cambridge Publications, 2012. He is co-recipient of the Best Paper Award of the Wireless Communications Symposium (WCS) in the IEEE International Conference on Communications (ICC'07), Glasgow, U.K., June 2007.

Dr. Karagiannidis has been a member of Technical Program Committees for several IEEE conferences such as ICC, GLOBECOM, VTC, etc. In the past he was Editor for Fading Channels and Diversity of the \textsc{IEEE Transactions on Communications}, Senior Editor of \textsc{IEEE Communications Letters} and Editor of the \textit{EURASIP Journal of Wireless Communications \& Networks}. He was Lead Guest Editor of the special issue on ``Optical wireless communications'' of the \textsc{IEEE Journal on Selected Areas in Communications} and Guest Editor of the special issue on ``Large-scale multiple antenna wireless systems''.

Since January 2012 he is the Editor-in Chief of \textsc{IEEE Communications Letters}.
\end{IEEEbiography}

\end{document}